\documentclass[twocolumn,prl,superscriptaddress]{revtex4}
\usepackage{stmaryrd}
\usepackage{amsmath}
\usepackage{amssymb}
\usepackage{graphicx}
\usepackage{dcolumn}
\usepackage{bm}
\usepackage{hyperref}
\usepackage{color}

\begin{document}


\title{Weak topological insulators induced by the inter-layer coupling: A first-principles study of stacked Bi$_2$TeI}

\author{Peizhe Tang}
\affiliation{Department of Physics and State Key Laboratory of Low-Dimensional Quantum Physics, Tsinghua University, Beijing 100084, People's Republic of China}

\author{Binghai Yan}
\email{yan@cpfs.mpg.de}
\affiliation{Max Planck Institute for Chemical Physics of Solids, D-01187 Dresden, Germany}
\affiliation{Max Planck Institute for the Physics of Complex Systems, D-01187 Dresden, Germany}

\author{Wendong Cao}
\affiliation{Department of Physics and State Key Laboratory of Low-Dimensional Quantum Physics, Tsinghua University, Beijing 100084, People's Republic of China}

\author{Shu-Chun Wu}
\affiliation{Max Planck Institute for Chemical Physics of Solids, D-01187 Dresden, Germany}

\author{Claudia Felser}
\affiliation{Max Planck Institute for Chemical Physics of Solids, D-01187 Dresden, Germany}

\author{Wenhui Duan}
\affiliation{Department of Physics and State Key Laboratory of Low-Dimensional Quantum Physics, Tsinghua University, Beijing 100084, People's Republic of China}
\affiliation{Institute for Advanced Study, Tsinghua University, Beijing 100084, China}

\date{\today}

\begin{abstract}
Based on first-principles calculations, we predict Bi$_2$TeI, a stoichiometric compound synthesized, to be a weak topological insulator (TI) in layered subvalent bismuth telluroiodides. Within a bulk energy gap of 80 meV, two Dirac-cone-like topological surface states exist on the side surface perpendicular to BiTeI layer plane. These Dirac cones are relatively isotropic due to the strong inter-layer coupling, distinguished from those of previously reported weak TI candidates. Moreover, with chemically stable cladding layers, the BiTeI-Bi$_{2}$-BiTeI sandwiched structure is a robust quantum spin Hall system, which can be obtained by simply cleaving the bulk Bi$_2$TeI.

\end{abstract}


\maketitle




Due to exotic electronic properties of the spin-helical Dirac fermion and the extensive application potential, topological insulators (TIs) have attracted a lot of attention in the last few years. TIs have an insulating state in the bulk and the gapless surface states (SS) at the boundary which are protected by the time-reversal symmetry (TRS). \cite{Hasan2010} For two dimensional (2D) TIs, the spin-filtered edge states propagate along the boundary with dissipationless spin and charge currents \cite{Bernevig2006,Konig2007}. In three dimension, TIs could be classified into strong and weak TIs according to the nature of SS. \cite{Fu2007_prb,Fu2007_prl,Moore2007} Strong TIs have odd number of Dirac cones on any surfaces, while weak TIs only exhibit topological non-trivial SS at the specific surfaces with even number of Dirac cones. The SS of weak TIs are also robust against strong perturbations that do not break TRS, showing the strong side of the topological properties. \cite{Imura2011,Ringel2012,Mong2012} And some novel topological quantum effects are proposed for the SS of weak TIs, such as half quantum spin Hall (QSH) effect \cite{Liu2011} and one-dimensional (1D) helical mode \cite{Imura2013,Ran2009}.

\begin{figure}[tbp]
\includegraphics[width=0.45\textwidth]{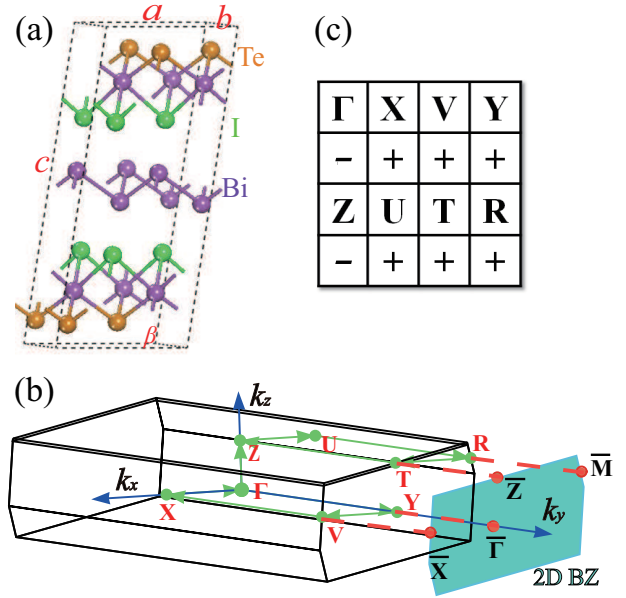}
\caption{\label{fig:fig1} (Color online) (a) The geometrical structure of 3D bulk Bi$_{2}$TeI, where the unit cell is indicated by the dashed lines. (b) The first Brillouin Zone (BZ) of 3D bulk Bi$_{2}$TeI and the projection to 2D BZ for the side surface which is labeled by cyan. (c) The product of the parity eigenvalues $\delta$ at the high symmetrical points which are marked in (b).}
\end{figure}

It is significant to locate real materials with these topological effects (i.e., weak TIs). \cite{Yan2012_review} Owing to the large spin-orbital coupling (SOC), the layered bismuth-based materials are the ideal candidates. As the prototypes of strong TIs, bismuth-chalcogenides such as Bi$_{2}$Se$_{3}$ and Bi$_{2}$Te$_{3}$, exhibit a typical spin-polarized Dirac cone at the $\Gamma$ point \cite{Zhang2009,Hsieh2009,Valla2012} and are thus ideal systems to study the magnetic properties when TRS is broken \cite{Chang2013,Zhang2013}. Furthermore, the bismuth bilayer is predicted to be 2D TI \cite{Murakami2006} and its topological edge states have been observed via scanning tunneling spectroscopy \cite{Yang2012}. For weak TIs, a traditional proposal is to build a three dimensional (3D) structure with weak interlay coupling via stacking the 2D TIs along the $z$ direction. In this proposal, the band inversions of stacked 3D compounds are completely contributed by QSH layers. Following this idea, it was predicted that KHgSb\cite{Yan2012} and Bi$_{14}$Rh$_{3}$I$_{9}$\cite{Rasche2013} with the honeycomb structures are weak TIs. In addition, weak TIs provide new possibilities to achieve 2D TIs by  cleaving off single QSH layers from the bulk, as commonly done for grapehene~\cite{Novoselov2005}, instead of fabricating complicated QWs \cite{Konig2007,Yang2012}.

\begin{figure*}[tbp]
\includegraphics[width=1.0\textwidth]{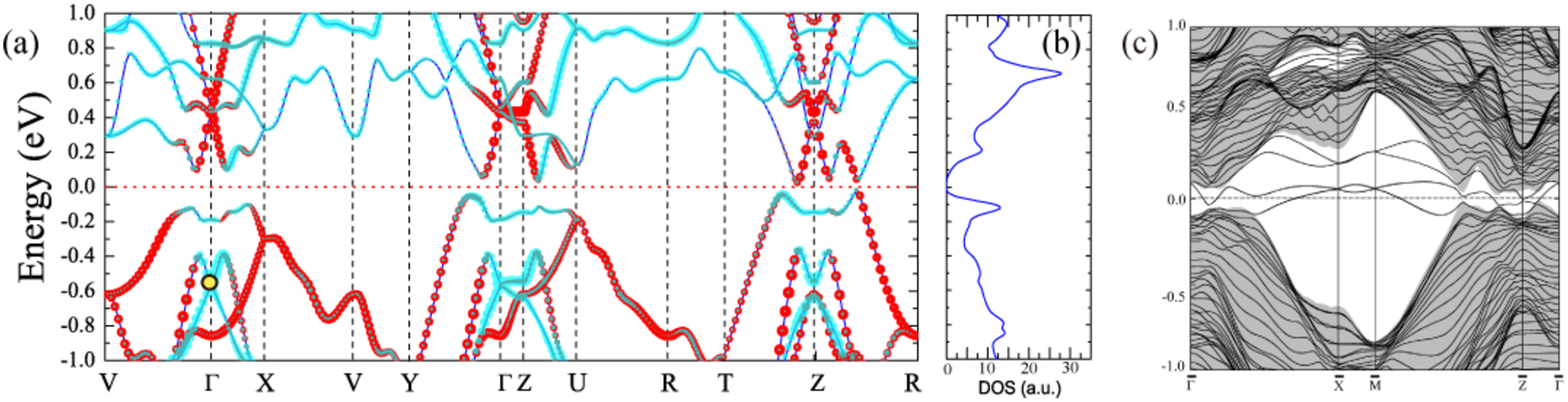}
\caption{\label{fig:fig2} (Color online) (a) Band structures along the high symmetry lines and (b) total density of states around the Fermi level of bulk Bi$_{2}$TeI. The red and cyan dots stand for the contribution from Bi bilayer and BiTeI layers separately. The Fermi level is set to zero (red dotted line). (c) The surface band structure along the high symmetry line in 2D BZ. The shadow indicates the projection of bulk states.}
\end{figure*}

In this letter, we study the topological and electronic properties of layered Bi$_2$TeI compounds which have been stoichiometrically synthesized recently \cite{Savilov2005,Isaeva2013}. Our \textit{ab initio} calculations show that the compounds are weak TIs with the band inversions occurring at the $\Gamma$ and Z points. However, different from the traditional proposal for weak TIs \cite{Yan2012,Rasche2013}, the band inversions are contributed by both the intercalated topologically trivial layers (BiTeI layers) and the topologically non-trivial layers (Bi bilayers). What is more, by changing the SOC strength artificially, we present a new band inversion picture to clarify the role of SOC during the band evolution process. In contrast to KHgSb \cite{Yan2012}, two isotropic spin-helical Dirac cones are shown at the $\overline{\rm M}$ and $\overline{\rm X}$ points in the Brillouin Zone (BZ) of the side surface. And from the surface formation energy (SFE) calculations, we find that the BiTeI-Bi$_{2}$-BiTeI sandwiched structure with Te terminations on the outer surface is stable and acts as a QSH insulator (i.e. a 2D TI).

The calculations were carried out by using density-functional theory (DFT) with the projector augmented wave method \cite{Blochl1999} and the generalized gradient approximation (GGA) with Perdew-Burke-Ernzerhof type functional \cite{Perdew1996}, as implemented in the Vienna \textit{ab initio} simulation package \cite{Kresse1996}. Plane wave basis set with a kinetic energy cutoff of 260 eV was used. The geometrical structure of Bi$_2$TeI is demonstrated in Fig. 1(a). It can be seen that Bi bilayer intercalates into two opposite BiTeI layers forming a monoclinic layered structure (space group \textit{C2/m}, $a=7.585${\AA}, $b=4.380${\AA}, $c=17.741${\AA}, $\beta$=98.20$^{\circ}$). Different from BiTeI compounds \cite{Bahramy2011,Ishizaka_many}, the inversion symmetry is preserved in Bi$_{2}$TeI. In the simulations, the lattice constants are taken from experimental observation \cite{Savilov2005,Isaeva2013} and all the inner atoms are fully relaxed until the residual forces are less than 0.01 eV/\AA. The Monkhorst-Pack \textit{k} points are 9$\times$11$\times$3 and SOC is included to calculate the electronic structure \cite{Hobbs2000}. For the surface formation energy (SFE) calculations \cite{note1}, a slab model is used with a vacuum layer larger than 10 {\AA} where the inversion symmetry is present, and the dipole correction is introduce to avoid spurious interactions between periodic images of the slabs \cite{Neugebauer1992}.

Figs. 2(a) and (b) respectively show the band structure along the high symmetry lines in the first BZ [as marked in Fig. 1(b)] and the total density of states for Bi$_2$TeI. It is found that bulk Bi$_2$TeI is a semiconductor with a direct band gap of about $80$ meV in the Z-R direction. Around the Fermi level, the valance and conduction bands are mainly contributed by the Bi bilayer and BiTeI layers respectively. But along the $\Gamma$-Z direction, the situation is reversed: the valence bands are mainly contributed by BiTeI layers through the strong hybridization between Bi $p_{x}$ and $p_{y}$ orbitals and Te $p_{z}$ orbital; and the conduction bands are mostly from the $p_{x}$ and $p_{y}$ orbital of Bi atoms in Bi bilayer. This fact suggests that the band inversions occur and this type of materials may be in a topological non-trivial phase. Similar to the BiTeI \cite{Bahramy2011,Ishizaka_many}, a band crossing can be observed in the bulk states [as marked by the yellow dot in Fig. 2(a)]. Due to the presence of the inversion symmetry, however, the crossing bands are doubly degenerate around the $\Gamma$ point and no Rashba-like splitting occurs here.

In order to clarify the topological properties, we use Fu and Kane's method \cite{Fu2007_prb} to calculate the $\mathbb{Z}_{2}$ invariants. In this method, $\mathbb{Z}_{2}$ ($\nu_{0},\nu_{1}\nu_{2}\nu_{3}$) of the 3D insulator could be determined from the product of the parity eigenvalues $\delta_{i}$ ($i=1,2,\ldots,8$) of the occupied states at the eight time reversal invariant momenta (TRIM) $\Gamma_{i}$ in the first BZ when the inversion symmetry is present. If an odd number of band inversions driven by SOC occur at the $\Gamma_{i}$ point and the parities of the inverted bands are different, $\delta_{i}$ is $-1$; otherwise, $\delta_{i}$ is $+1$. The main topological index $\nu_{0}$ is determined by $(-1)^{\nu_{0}}=\prod_{i=1}^{8}\delta_{i}$. The other three weak topological indices $\nu_{k}$ ($k=1,2,3$) are determined by $(-1)^{\nu_{k}}=\prod_{i=1}^{4}\delta_{i}$, for which $\Gamma_{i}$ reside in the same plane. For Bi$_{2}$TeI, the eight TRIM points are $\Gamma$ (0,0,0), $\rm X$ (0.5, 0, 0), $\rm V$ (0.5, 0.5, 0), $\rm Y$ (0, 0.5, 0), $\rm Z$ (0, 0, 0.5), $\rm U$ (0.5, 0, 0.5), $\rm T$ (0.5, 0.5, 0.5) and $\rm R$ (0, 0.5, 0.5) [see Fig. 1(b)]. Thus four topological indices are determined as $(-1)^{\nu_{0}}=\delta_{\Gamma}\delta_{Z}\delta_{X}\delta_{U}\delta_{V}\delta_{T}\delta_{Y}\delta_{R}$, $(-1)^{\nu_{1}}=\delta_{\Gamma}\delta_{Z}\delta_{X}\delta_{U}$, $(-1)^{\nu_{2}}=\delta_{\Gamma}\delta_{Z}\delta_{T}\delta_{Y}$ and $(-1)^{\nu_{3}}=\delta_{\Gamma}\delta_{X}\delta_{V}\delta_{Y}$. According to the calculated products of the parity eigenvalues ($\delta_{i}$) [as shown in Fig. 1(c)],  the bulk Bi$_2$TeI belongs to the $\mathbb{Z}_{2}$ ($0,001$) class of weak TIs.

To demonstrate the topological non-trivial SS, we have further calculated the surface band structure  using maximally localized Wannier functions \cite{Marzari1997} extracted from our \textit{ab initio} results. The tight-binding slab model containing more than $10$ unit-cell-thick layers along the $b$ direction [shown in Fig. 1(a)] is used, where the side surface is selected as the \textit{ac} plane of Bi$_{2}$TeI lattice and the Bi bilayer exhibits an armchair type of termination on the boundary. The 2D BZ is shown in Fig. 1(b) marked by cyan. As shown in Fig. 2 (c), the SS is plotted as solid lines and the shaded regions denote the energy spectrum from bulk. Around the Fermi level, the metallic middle-gap states are contributed by the side surface: they emerge from the conduction bands, cross at two TRIM points and enter the valance bands, exhibiting the topological non-trivial properties of weak TIs. What is more, two Dirac cones are isotropic and locate at the $\overline{\rm X}$ and $\overline{\rm M}$ points which are not the projections of bulk $\Gamma$ and Z points to the side surface. This character is in contrast to that of KHgSb, but similar to the property of bismuth bilayer (a 2D TI) where the crossing of edge states is at $\overline{\rm M}$ point instead of $\overline{\Gamma}$ point \cite{Murakami2006,Yang2012}.

\begin{figure}[tbp]
\includegraphics[width=0.45\textwidth]{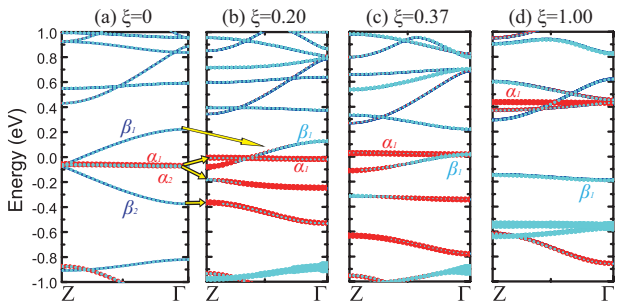}
\caption{\label{fig:fig3} (Color online) Band structures along the Z-$\Gamma$ direction with different spin-orbital coupling ratios ($\xi$) for bulk Bi$_{2}$TeI: (a) $\xi=0$, (b) $\xi=0.20$, (c) $\xi=0.37$ and (d) $\xi=1.00$. Red and cyan dots stand for the contributions from Bi bilayer and BiTeI layers respectively, and the Fermi level is set to zero.}
\end{figure}

Generally speaking, for the bismuth-based TIs the band inversions are driven by the large SOC. Thus, to observe the band evolution with the change of SOC ratio ($\xi=\lambda/\lambda_{0}$) is an effective method to clarify the origin of band inversion, where $\lambda_{0}$ and $\lambda$ are respectively the original SOC strength and the SOC strength used in the DFT calculations. Fig. 3 shows the calculated band structures of bulk Bi$_{2}$TeI along the Z-$\Gamma$ direction with different $\xi$. For the bands without SOC ($\xi=0$), two states mainly from the Bi bilayer ($\alpha_{1}$ and $\alpha_{2}$) are degenerate and fully occupied; while the other bands ($\beta_{1}$ and $\beta_{2}$) are mainly contributed by BiTeI layers. And $\beta_{1}$ state is partially occupied, indicating that Bi$_{2}$TeI bulk should behave like a normal metal if SOC is not included. With increasing $\xi$, as shown in Fig. 3(b) and marked by yellow arrows, the hybridization between $\alpha_{1}$ and $\alpha_{2}$ ($\beta_{1}$ and $\beta_{2}$) is changed, and thus the band crossing at the Z point disappears. The state contributed by Bi bilayer ($\alpha_{1}$) crosses with the other from BiTeI layers ($\beta_{1}$) around the Fermi level. Owing to the splitting, the occupied states at the Z point exhibit the topological non-trivial property ($\delta_{Z}$ is $-1$), but $\delta_{\Gamma}$ is still $+1$. Further increasing $\xi$, we can observe a topological phase transition at the critical ratio ($\xi=0.37$) [shown in Fig. 3(c)]: the bands ($\alpha_{1}$ and $\beta_{1}$) cross each other at the $\Gamma$ point, similar to a 3D Dirac cone. Above the critical ratio, $\alpha_{1}$ state is unoccupied and $\beta_{1}$ state is fully occupied [see Fig. 3(d)], indicating that the system becomes a weak TI ($\delta_{Z}=-1$ and $\delta_{\Gamma}=-1$).

In the previous proposal of weak TIs \cite{Yan2012,Rasche2013}, the band inversion occurs inside each single QSH layer, while the role of the topologically trivial intercalated layers is just to weakly couple neighboring QSH layers. In Bi$_{2}$TeI, however, a very different band inversion scenario is observed: the inversion is contributed by both the topologically non-trivial Bi bilayer and the trivial BiTeI layers. During the inversion process, the role of SOC is different at the Z and $\Gamma$ points: SOC splits the degenerate states at the Z point (i.e., opening a band gap there), and simultaneously induces the band inversion at the $\Gamma$ point, similar to the case of Bi$_{2}$Se$_{3}$\cite{Zhang2009,Yan2012_review}. Evidently, the band inversion picture of Bi$_{2}$TeI is unique and this fact could improve our understanding of the band inversion mechanism in weak TIs and open a way for the search of new TI materials.

Due to its layered structure, a 2D sheet can possibly be cleaved from the bulk Bi$_{2}$TeI. Our SFE calculations within local-density approximation show that a SFE of 9 meV/{\AA}$^{2}$ is required to cleave a BiTeI-Bi$_{2}$-BiTeI sandwiched structure, which is comparable to the formation energy of a single layer of graphene (5 meV/{\AA}$^{2}$) or MoS$_2$ (6 meV/{\AA}$^{2}$). Compared to a Bi$_{2}$-BiTeI-BiTeI-Bi$_{2}$ type of cleaved sheet (with a SFE of 18 meV/{\AA}$^{2}$), it is  energetically more stable, and the same trend has been confirmed by GGA and van der Waals density functional calculations \cite{Dion2004}. The stable structure exhibits an energy gap of 40 meV, slightly different from its 3D counterpart. It is interesting to find that the exfoliated layer can be a QSH insulator. To elucidate this finding, the topological edge states of an 1D armchair BiTeI-Bi$_{2}$-BiTeI nanoribbon are calculated using the maximally-localized Wannier functions \cite{Marzari1997} from \textit{ab inito} calculations. As shown in Fig. 4, the 1D edge states can be observed inside the bulk gap and the Dirac-type of band crossing is found at the $\Gamma$ point of the 1D BZ. Although trivial and nontrivial edge states co-exist, a generic Fermi level intersects the edge state three times between the $\Gamma$ and $\rm X$ points, confirming the topologically nontrivial feature of the system. It should be pointed out that edge states of the BiTeI-Bi$_{2}$-BiTeI film behave differently from those of isolated Bi bilayers, where Dirac-point crossing exists at the Brillouin zone boundary \cite{Murakami2006,Yang2012}. Moreover, we note that the BiTeI layer is chemically more inactive than the Bi$_2$ layer and thus protect the Bi$_2$ layer from the ambient environment.

\begin{figure}[tbp]
\includegraphics[width=0.45\textwidth]{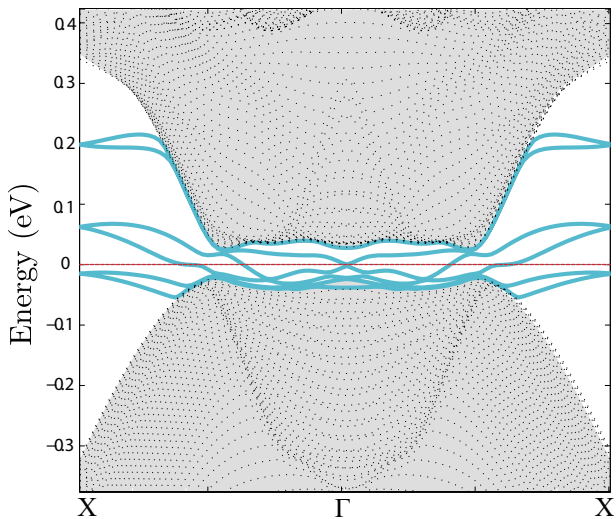}
\caption{\label{fig:fig4} (Color online) The edge states of a BiTeI-Bi$_{2}$-BiTeI ribbon along the $a$ direction. The blue lines represent the metallic edge states and gray shadow stands for the 2D bulk states. The Fermi level is set as zero and marked by the red dotted line.}
\end{figure}

In conclusions, We found that Bi$_{2}$TeI belongs to $\mathbb{Z}_{2}$ ($0;001$) class of weak TIs with two relatively isotropic Dirac cones on the side surface. In contrast to previously proposed weak TIs, a new band inversion mechanism is demonstrated in which the inter-layer coupling between the QSH layers and trivial layers induces the band inversions at both $\Gamma$ and Z points. Moreover, the exfoliated thin film of Bi$_{2}$TeI can act as a QSH insulator with helical edge states. The layered Bi$_2$TeI is a prototype of weak TIs and also provides a new platform to realize the QSH effect.

We acknowledge the support of the Ministry of Science and Technology of China (Grant Nos. 2011CB921901 and 2011CB606405), and the National Natural Science Foundation of China (Grant Nos. 11074139). B.Y. and C.F. acknowledges financial support from the European Research Council Advanced Grant (ERC 291472).


\begin{references}
\bibitem{note1} The surface formation energy is defined as $\frac{1}{2A}[E_{\rm sur}-N_{\rm Bi}\mu_{\rm Bi}-N_{\rm Te}\mu_{\rm Te}-N_{\rm I}\mu_{\rm I}]$, where $A$ is the surface area and $E_{\rm sur}$ is the total energies of 2D thin film. $\mu_{\rm Bi}$, $\mu_{\rm Te}$ and $\mu_{\rm I}$ are the chemical potentials of Bi, Te and I atom respectively. $N_{\rm Bi}$, $N_{\rm Te}$ and $N_{\rm I}$ are the numbers of Bi, Te and I atoms respectively in the structure. Note that $2\mu_{\rm Bi}+\mu_{\rm Te}+\mu_{\rm I}$ is equal to the total energy per formula of bulk Bi$_{2}$TeI.
\end{references}
\end{document}